\documentclass[prl,twocolumn,nofootinbib, preprintnumbers, superscriptaddress]{revtex4}

\usepackage{amsmath, amssymb, slashed, braket}
\usepackage{graphicx}
\usepackage{epstopdf}
\usepackage{float,appendix}
\usepackage[colorlinks=true,
            linkcolor=blue,
            urlcolor=blue,
            citecolor=green,          
            bookmarks=true,
            bookmarksnumbered=true,
            breaklinks=true,
            pdfpagemode=FullScreen,
            pdfstartview=FitBH]{hyperref}
\usepackage{esint}
\usepackage[normalem]{ulem}
\usepackage{siunitx}

\usepackage{color}

\definecolor{Orange}{cmyk}{0,0.61,0.87,0}
\definecolor{JungleGreen}{cmyk}{0.99,0,0.52,0}
\definecolor{OliveGreen}{cmyk}{0.64,0,0.95,0.40}
\definecolor{Brown}{cmyk}{0,0.81,1,0.60}
\definecolor{RoyalBlue}{cmyk}{0.71,0.53,0,0.12}
\definecolor{Gray}{cmyk}{0,0,0,0.40}
\definecolor{LightPink}{cmyk}{0.0,0.25,0,0}
\definecolor{LLightPink}{cmyk}{0.0,0.10,0,0}
\definecolor{LightBlue}{cmyk}{0.25,0,0,0}
\definecolor{LightGray}{cmyk}{0,0,0,0.2}

\usepackage{xcolor}
\definecolor{gesfpurple}{rgb}{0.47,0.19,0.42}

\definecolor{gesflanse}{rgb}{0.00,0.50,0.50}

\definecolor{gesfblue}{rgb}{0.08,0.42,0.76}

\definecolor{gesfred}{rgb}{1,0,0}

\definecolor{gesfwhite}{rgb}{1,1,1}

\definecolor{gesfblack}{rgb}{0,0,0}

\newcommand{\geqn}[1]{Eq.\,\hypersetup{linkcolor=blue}(\ref{#1})\hypersetup{linkcolor=blue}}
\newcommand{\gfig}[1]{{\hypersetup{linkcolor=violet}Fig.\,\ref{#1}\hypersetup{linkcolor=blue}}}

\newcommand\redout{\bgroup\markoverwith{\textcolor{red}{\rule[.5ex]{2pt}{0.4pt}}}\ULon}

\begin{document}

\title{
    Detecting meV-Scale Dark Matter \\ via \\ Coherent Scattering with an Asymmetric Torsion Balance
}


\author{}
\affiliation{Tsung-Dao Lee Institute \& School of Physics and Astronomy, Shanghai Jiao Tong University, China}
\affiliation{College of Science, China University of Petroleum (East China), Qingdao 266580, China}
\affiliation{Kavli IPMU (WPI), UTIAS, University of Tokyo, Kashiwa, 277-8583, Japan}
\affiliation{National Gravitation Laboratory, MOE Key Laboratory of Fundamental Physical Quantities Measurement \& Hubei Key Laboratory of Gravitation and Quantum Physics, School of Physics, Huazhong University of Science and Technology, Wuhan 430074, People's Republic of China}
\affiliation{State Key Laboratory of Dark Matter Physics, Tsung-Dao Lee Institute \& School of Physics and Astronomy, Shanghai Jiao Tong University, Shanghai 200240, China}
\affiliation{Key Laboratory for Particle Astrophysics and   Cosmology (MOE) \& Shanghai Key Laboratory for Particle Physics and Cosmology, Shanghai Jiao Tong University, Shanghai 200240, China}

\author{Pengshun Luo} 
\affiliation{National Gravitation Laboratory, MOE Key Laboratory of Fundamental Physical Quantities Measurement \& Hubei Key Laboratory of Gravitation and Quantum Physics, School of Physics, Huazhong University of Science and Technology, Wuhan 430074, People's Republic of China}

\author{Shigeki Matsumoto} \email{shigeki.matsumoto@ipmu.jp}
\affiliation{Kavli IPMU (WPI), UTIAS, University of Tokyo, Kashiwa, 277-8583, Japan}

\author{Jie Sheng}
\email{shengjie04@sjtu.edu.cn}
\affiliation{Tsung-Dao Lee Institute \& School of Physics and Astronomy, Shanghai Jiao Tong University, China}
\affiliation{State Key Laboratory of Dark Matter Physics, Tsung-Dao Lee Institute \& School of Physics and Astronomy, Shanghai Jiao Tong University, Shanghai 200240, China}
\affiliation{Key Laboratory for Particle Astrophysics and   Cosmology (MOE) \& Shanghai Key Laboratory for Particle Physics and Cosmology, Shanghai Jiao Tong University, Shanghai 200240, China}

\author{Chuan-Yang Xing}
\email{cyxing@upc.edu.cn}
\affiliation{Tsung-Dao Lee Institute \& School of Physics and Astronomy, Shanghai Jiao Tong University, China}
\affiliation{College of Science, China University of Petroleum (East China), Qingdao 266580, China}
\affiliation{Key Laboratory for Particle Astrophysics and   Cosmology (MOE) \& Shanghai Key Laboratory for Particle Physics and Cosmology, Shanghai Jiao Tong University, Shanghai 200240, China}

\author{Lin Zhu} \email{zhulin36@mail.hust.edu.cn}
    \affiliation{National Gravitation Laboratory, MOE Key Laboratory of Fundamental Physical Quantities Measurement \& Hubei Key Laboratory of Gravitation and Quantum Physics, School of Physics, Huazhong University of Science and Technology, Wuhan 430074, People's Republic of China}

\author{Zhi-Jie Zhuge}
    \affiliation{National Gravitation Laboratory, MOE Key Laboratory of Fundamental Physical Quantities Measurement \& Hubei Key Laboratory of Gravitation and Quantum Physics, School of Physics, Huazhong University of Science and Technology, Wuhan 430074, People's Republic of China}

\begin{abstract}

Dark matter with mass in the crossover range between wave dark matter and particle dark matter, around $(10^{-3},\, 10^3)$\,eV, remains relatively unexplored by terrestrial experiments. In this mass regime, dark matter scatters coherently with macroscopic objects. The effect of the coherent scattering greatly enhances the accelerations of the targets caused from dark matter collisions by a factor of $\sim 10^{23}$. We propose a novel torsion balance experiment with test bodies of different geometric sizes to detect such dark matter-induced acceleration. This method provides the strongest sensitivity on the scattering cross-section between the dark matter and a nucleon in the mass range of $(10^{-3}, 1)\,$eV.

\end{abstract}

\maketitle 

\noindent
{\bf Introduction} -- More than 80 percent of the matter in our Universe today is dark matter (DM)\,\cite{Planck:2018vyg}. However, the nature of DM, neither its mass nor interaction with standard model (SM) particles, is unknown\,\cite{Bertone:2004pz, Young:2016ala, Arbey:2021gdg}. For a long time, the Weakly Interacting Massive Particle has been the mainstream candidate\,\cite{Lee:1977ua}. DM direct detection experiments based on the scattering between the DM and a nucleon have primarily targeted it\,\cite{Lin:2019uvt, Cooley:2021rws}. These, however, have not yielded positive results so far, imposing strong constraints on DM heavier than $\mathcal{O}(10)$\,GeV\,\cite{PandaX-4T:2021bab, LZ:2022lsv, XENON:2023cxc}.

There is broader mass speculation regarding the DM: particle DM with masses ranging from $\mathcal{O}(1)$\,eV to the Planck mass and wave DM with masses ranging from $\mathcal{O}(10^{-22})$\,eV to $\mathcal{O}(1)$\,eV\,\cite{Billard:2021uyg}. DM candidates also include macroscopic objects, e.g., the primordial black hole\,\cite{Carr:2020xqk, Carr:2021bzv}. Diverse direct detection methods have emerged to cover this wide range of masses and interactions.

In addition to the experiments relying on the DM scattering off a nucleon for the DM with masses above GeV, those on the scattering off an electron\,\cite{Essig:2011nj} and off crystals to induce phonon excitation\,\cite{Knapen:2017ekk, Griffin:2018bjn, Coskuner:2021qxo, Campbell-Deem:2022fqm} are utilized to detect the particle DM in the keV to MeV mass range. On the other hand, for the ultra-light wave DM with mass smaller than $\mathcal{O}(10^{-6})$\,eV, interferometers\,\cite{Stadnik:2014tta, Buchmueller:2023nll} and atomic clocks\,\cite{Filzinger:2023qqh} can detect its influence on fundamental constants. The haloscope\,\cite{ADMX:2019uok, QUAX:2020adt, CAPP:2020utb, HAYSTAC:2023cam}, and helioscope\,\cite{CAST:2017uph, IAXO:2019mpb} are sensitive to axion-like interactions. 
For the DM with masses in the crossover region between the wave DM and the particle DM, the $(10^{-5},\,10^3)$\,eV range, especially those with $Z_2$ symmetry, there are currently no particularly sensitive detection methods available.

In this paper, we propose a special torsion balance experiment with test bodies of different sizes to sensitively detect the DM in the above mass range for the first time. Although the energy transfer of such a DM is insufficient to detect, their continuous momentum transfer can induce acceleration on macro-objects\,\cite{Domcke:2017aqj, Graham:2015ifn, Carney:2019cio, Day:2023mkb}.\footnote{
    In early studies, similar ideas were used for cosmic neutrino background (C$\nu$B) detection\,\cite{zel1981neutrino,shvartsman1982possibility, Ringwald:2004np, Domcke:2017aqj, Akhmedov:2018wlf, Shergold:2021evs}.
    Unfortunately, due to the diffuse flux of cosmic neutrinos and their small scattering cross-section with nucleons, the resulting acceleration of the objects is far below the detectability threshold of current technologies.}
However, usual torsion balance experiments with the test masses sharing the same mass and outside dimensions cannot detect DM-induced accelerations, as forces from the DM scattering on test masses are the same. We show an efficient configuration for detecting the DM: 
the test bodies, in the form of solid cubes and hollow cubical shells,
have the same mass but are vastly different in size. 
Due to the long de Broglie wavelength of light particles, the scattering cross-section is greatly enhanced through the coherent effect\,\cite{Fukuda:2018omk, Fukuda:2021drn}, resulting in the most stringent sensitivity on the DM scattering off a nucleon in the mass range of ($10^{-3}$,\,$1$)\,eV. We use natural units with $c = \hbar = 1 $ throughout the text.

\noindent
{\bf DM Coherent Scattering with Cube \& Shell} -- Due to the rotation of the solar system around the Galactic center, there is a relative velocity between the solar system and DM halo, typically the DM velocity in the laboratory frame, $v_\chi \simeq 10^{-3}$. Assuming the DM has an interaction with a nucleon $N = n, p$ with a scattering cross-section $\sigma_{\chi N}$, DM collisions would impact a force to a target on Earth, resulting in its acceleration,
\begin{equation}
    a \sim
    \frac{1}{m_{\mathrm{tot}}}
    \frac{\rho_\chi}{m_\chi}
    \sigma_{\mathrm{tot}} v_\chi\,q,
\label{acc_chi_N}
\end{equation}
where $m_{\mathrm{tot}}$ is the total mass of the target, $\sigma_{\mathrm{tot}}$ is the total scattering cross-section between the DM and the target, and $\rho_\chi \simeq 0.4 \,\mathrm{GeV} / \mathrm{cm}^3$ is the local DM density. If the DM mass $m_\chi$ is much smaller than the nucleon mass $m_N$, the DM is bounced away with almost the same momentum magnitude. The momentum transfer $q$ of the scattering is roughly the incident DM momentum, $q \simeq m_\chi v_\chi$. Then, the DM-induced acceleration is proportional to the ratio of the cross-section and the target mass, $a \propto \sigma_{\mathrm{tot}}/m_{\mathrm{tot}}$, independent of the target size if both quantities are proportional to the nucleon number in the target. This is usually the case for the DM with a large mass or, equivalently, with a short de Broglie wavelength, which generally causes too small acceleration.

However, when the inverse of momentum transfer $1/q$
is larger than the size of the target, the total cross-section is enhanced quadratically concerning the nucleon number by the coherent scattering effect. For example, the coherent DM-nucleus scattering cross-section $\sigma_{\chi A} = A^2 \sigma_{\chi N}$ has an enhancement of $A^2$ compared to the DM-nucleon scattering where $A$ is the atomic number of the nucleus. This effect has already been confirmed in the Coherent Elastic Neutrino-Nucleus Scattering (CE$\nu$NS)\,\cite{COHERENT:2017ipa}. Similarly, when the wavelength of the DM is larger than the size of a macroscopic object, the DM coherently scatters off the entire object, which contains $N_A$ atoms of the order of Avogadro's number, $N_A \sim 10^{23}$. Then, the total cross-section $\sigma_{\mathrm{tot}}$ is further enhanced by a factor of $N_A^2$ compared to $\sigma_{\chi A}$. Therefore, according to \geqn{acc_chi_N}, the acceleration would be enhanced by a factor of $N_A$. It greatly improves the detection capability for a weakly interacting DM with mass $m_\chi \lesssim 10^{-1}$\,eV, corresponding to the de Broglie wavelength of $1 / (m_\chi v_\chi) \gtrsim \mathcal{O}(1)$\,mm.

The coherent scattering of the DM with a macroscopic object originates from the wave nature of the DM. In quantum mechanics, the scattering of the DM particle with a nucleus can be described as a plane wave, which is partially transformed into a weak spherical wave radiating from the nucleus potential as 
$
    \Psi_\mathrm{sc} = f(\mathbf{k},\mathbf{k}') e^{i k r}/r
$\,\cite{boothroyd:2020principles}.
Here, $\mathbf{k}$ and $\mathbf{k}'$ are the momenta of the incident and scattered DM particle, respectively, with magnitude $k \equiv |\mathbf{k}| = |\mathbf{k}'|$, and $r$ is the distance from the nucleus $r \equiv |\mathbf{r}|$. The coefficient $f(\mathbf{k},\mathbf{k}')$ is the scattering amplitude. In the case of the DM scattering with multiple nuclei, the DM plane wave interacts with the potential of each nucleus at different locations, emitting a spherical wave with the same scattering amplitude but different phases due to their differences in propagation distance. The total scattered wave is a superposition as\,\cite{boothroyd:2020principles},
\begin{equation}
    \Psi_\mathrm{sc} =
    f(\mathbf{k},\mathbf{k}') \frac{e^{i k r}}{r} \sum_{i=1}^{N_A} e^{i \mathbf{q} \cdot \mathbf{r}_i}.
    \label{scattering_wave_function}
\end{equation}
Here, each nuclear position is labeled by $\mathbf{r}_i$, and the momentum transfer is denoted by $\mathbf{q} \equiv \mathbf{k} - \mathbf{k}'$. At the cross-section level, the extra factor becomes, 
\begin{equation}
    \sigma_{\chi A} \rightarrow \sigma_{\mathrm{tot}} = \sum_{i,j}^{N_A}  e^{i \mathbf{q} \cdot \Delta {\bf r}_{ij} }  \sigma_{\chi A},
\end{equation}
with $\Delta {\bf r}_{ij} \equiv \mathbf{r}_i - \mathbf{r}_j$ being the distance between two nuclei.
Once the inverse of the momentum transfer amplitude $q \equiv |{\bf q}|$ is much larger than the spatial extent of these nuclei, i.e., $1/q \gg \Delta {\bf r}_{ij} $, $e^{i \mathbf{q} \cdot \Delta {\bf r}_{ij} } \simeq 1$. Then, the total cross-section has the enhancement factor of $N_A^2$. On the other hand, if $1/q \ll \Delta {\bf r}_{ij}$, all the phase factors cancel each other except for terms $i = j$ with $|\Delta {\bf r}_{ii}| = 0$; it only gives an enhancement of $N_A$ in the cross-section.
With the above limiting cases, the summation of the phase factors for any $\mathbf{q}$ can be parameterized as 
\begin{equation}
    \sum_{i, j=1}^{N_A}
    e^{ i \mathbf{q} \cdot \Delta \mathbf{r}_{ij} } =
    N_A + \left( N_A^2 - N_A \right) |F({\bf q})|^2 
    \label{phase_factor_sum}
\end{equation}
with a form factor $|F({\bf q})|^2 \equiv 1/(N_A^2 - N_A) \times  \sum_{i \neq j} e^{ i \mathbf{q} \cdot \Delta {\bf r}_{ij} }$. In the limit of a large $N_A$, the summation for the factor can be replaced by an integration in the continuous limit,
\begin{equation}
    |F({\bf q})|^2 
= 
	\frac{1}{V^2}
	\int d^3 \mathbf{r}_i\,d^3 \mathbf{r}_j\,e^{ i \mathbf{q} \cdot \Delta {\bf r}_{ij} } .
\end{equation}

\begin{figure}[t]
    \centering
    \includegraphics[width=8.5cm]{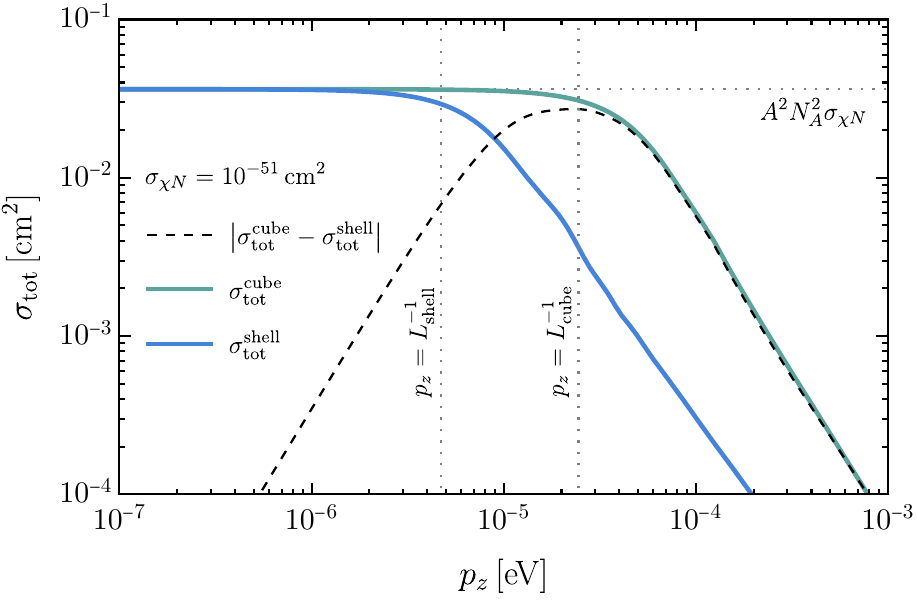}
    \caption{
    The total scattering cross-sections of the DM on a cube (solid green curve) and a shell (solid blue curve) as a function of the DM momentum along a rotation axis, $p_z$. 
    Their difference is also shown as a dashed black curve.
    The DM-nucleon scattering cross-section is $\sigma_{\chi N} = 10^{-51}\,$cm$^2$. The cube and the shell sizes are given in TABLE~\ref{tab:parameter}.}%
    \label{fig:xsec}
\end{figure}

As discussed later, our proposed torsion balance consists of a cube and a cubical shell with equal masses but different edge lengths. Thus, we use them as examples to calculate the coherent form factor. For a cube with an edge length $L$, the above integral can be performed to obtain,
\begin{equation}
    F_{\mathrm{cube}} (\mathbf{q}, L)
=
    \frac{e^{i q_x L}-1}{i q_x L}
    \frac{e^{i q_y L}-1}{i q_y L}
    \frac{e^{i q_z L}-1}{i q_z L} ,
\end{equation}
where $q_{x,y,z}$ are the momentum components along the corresponding directions.
In the $q L \ll 1$ limit, the form factor becomes one, as expected, and the enhancement factor in Eq.\,\eqref{phase_factor_sum} scales as $N_A^2$. The form factor vanishes for the $q L \gg 1$ limit, and the enhancement factor returns to $N_A$.
In the intermediate region, $|F_{\mathrm{cube}}(\mathbf{q}, L)|^2$ starts to decrease at around $1/q \simeq L$, and the coherent effect weakens.
Such a feature holds for other geometries. For a cubical shell with outer edge length $L_1$ and inner edge length $L_2$, the form factor,
\begin{align}
    F_{\mathrm{shell}}(\mathbf{q}, L_1, L_2) =
    \frac{L_1^3
    F_{\mathrm{cube}} (\mathbf{q}, L_1) -
    L_2^3
    F_{\mathrm{cube}} (\mathbf{q}, L_2)}{L_1^3 - L_2^3} ,
\end{align}
starts decreasing at $1/q \sim L_1 \equiv L_{\mathrm{shell}}$, as seen in \gfig{fig:xsec}.

The total differential cross-section between the DM and a macroscopic object is given by
\begin{equation}
\frac{d\sigma_\mathrm{tot}}{d {\bf q}} = A^2 \left[N_A+\left(N_A^2-N_A\right) |F({\bf q})|^2\right] \frac{d\sigma_{\chi N}}{d {\bf q}} .
\label{dsigmaA2}
\end{equation}
The \geqn{dsigmaA2} above is applicable to most scenarios where DM and nucleons interact weakly.
When the interaction between the DM and a nucleon is strong, most DM wave is scattered away at the surface of the object, and \geqn{scattering_wave_function} no longer works. Therefore, in the strong interaction limit, the total cross-section saturates to the geometrical size $S$ of the target\,\cite{Fukuda:2018omk, Fukuda:2021drn}. We note that the DM parameter space focused in this letter fall within the domain of weak interactions.

The total scattering cross-section, assuming an isotropic DM-nucleon scattering, is illustrated in \gfig{fig:xsec}.
The dimensions of the cube and the shell are provided in TABLE~\ref{tab:parameter}.
At small momentum transfer, the total cross-section is constant as 
$A^2 N_A^2 \sigma_{\chi N}$, and it rapidly decreases when $q \gtrsim 1/L_{\mathrm{cube,\,shell}}$.
\vspace{0.3cm}

\noindent
{\bf Proposed Experimental Setup \& Backgrounds} -- 
Currently, the torsion balance is a highly sensitive instrument used to measure acceleration difference between two types of test bodies. Its sensitivity can reach $\delta a \sim 10^{-13}$\,cm/s$^2$\,\cite{Schlamminger:2007ht,Wagner:2012ui, Luo:2018EP}. However, the conventional torsion balances are ineffective at detecting coherent scattering induced by DM, as the test bodies are typically designed with identical external dimensions. As a result, the test bodies experience nearly identical forces from the DM wind. Consequently, the net torque is canceled out, rendering the system  insensitive to DM scattering.

\begin{figure}[bt]
		\centering
		\includegraphics[width=8.5cm]{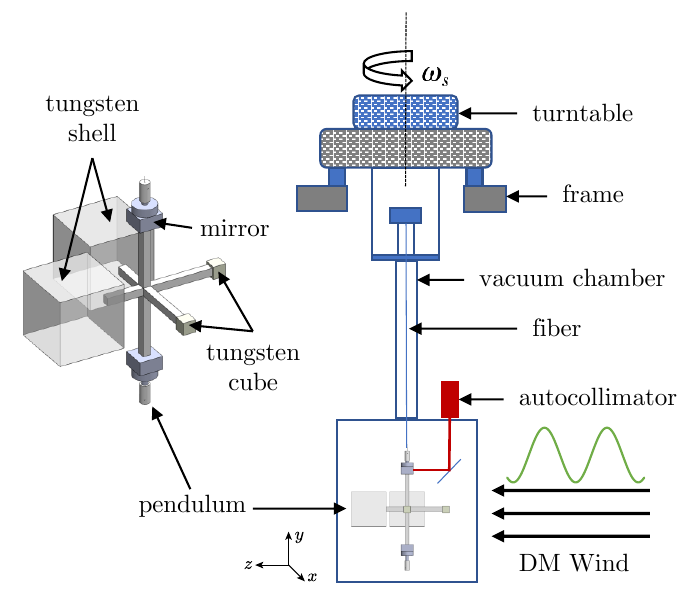}
		\caption{Schematic picture of the torsion balance experiment to detect the DM. The masses of test bodies are the same.}
		\label{fig:Torsion Balance}
	\end{figure}

To make the system responsive to DM-induced torque, we propose a specially designed torsion balance featuring test bodies of different sizes, as shown in \gfig{fig:Torsion Balance}. For simplicity, the test bodies are assumed to be made of the same material such as tungsten.\footnote{
	Using the same material (atom) benefits us in interpreting the measurement result accurately. In contrast, using different materials (atoms) would cause a systematic uncertainty associated with the material(atom)-dependent scattering cross-sections.} The proposed design is similar to that previously developed for equivalence principle test\cite{Luo:2018EP}. The pendulum, with a total mass of approximately $72.4\,\mathrm{g}$, carries four test bodies: two solid cubes with side length of $0.804\,\mathrm{cm}$ and two hollow shells with thickness of $\delta =50\,\mu\mathrm{m}$ and edge length of $4.166\,\mathrm{cm}$ (see TABLE~\ref{tab:parameter}). Each test body is positioned approximately $4.652\,\mathrm{cm}$ from the pendulum's center mass. The test bodies are glued to five glass rectangular rods, along with two glass mirrors. Owing to their different sizes, the solid cubes and hollow shells experience different scattering by DM, resulting in a torque on the pendulum along the direction of the suspension fiber. The angular position of the pendulum is monitored using an autocollimator. To enable precise detection of the DM-induced torque, the pendulum is placed in a vacuum chamber and continuously rotated by a turntable, together with the autocollimator, causing the torque to manifest as a sinusoidal signal. The rotation frequency can be tuned to optimize the signal-to-noise ratio of the torsion balance.

	To estimate the detection sensitivity to DM scattering, several systematic effects are taken into account. These include magnetic coupling \cite{Ding:2025dim}, local gravitational coupling, temperature gradients, deflection of the rotary axis from the local vertical \cite{zhao_influence_2021}, and variations in rotary rate, all of which were carefully evaluated in a previous equivalence-principle test \cite{Luo:2018EP}. 
    In the case of local gravitational coupling, the asymmetry in the pendulum's mass distribution, characterized by multipole moments $q_{l1}\, (l=2,3,\cdots)$ \cite{gong_high-order_2024}, is larger than in previous test. 
    Notably, the magnitude of the moment is approximately $|q_{31}|\approx 10\,\mathrm{g \, cm^3}$, which is ten times larger than that in the previous setup. This dominant gravitational coupling increases the associated uncertainty to $7.9\times 10^{-13}\,\mathrm{cm/s^2}$. 
    Due to the differing sizes of the test bodies, the pendulum's sensitivity to the temperature gradients is expected to be amplified. A conservative amplification factor of 10 is adopted, yielding an associated uncertainty of $9.4\times 10^{-13}\,\mathrm{cm/s^2}$. The uncertainties from other systematic effects are estimated to remain unchanged from the previous test, at $4.1\times 10^{-13}\,\mathrm{cm/s^2}$. The statistical uncertainty is also assumed to remain unchanged, as it is primarily limited by the internal friction of the suspension fiber, which remains the same as in the previous test, $4.1\times 10^{-13}\,\mathrm{cm/s^2}$. Therefore, the total uncertainty in the measured acceleration difference between the tungsten cube and shell is estimated to be approximately $ \delta a = 1.4\times 10^{-12}\,\mathrm{cm/s^{2}}$ at a $1 \sigma$ confidence level. This corresponds to $|\delta a| < 2.7\times 10^{-12}\,\mathrm{cm/s^{2}}$ at \SI{95}{\percent} confidence level.

\begin{table}[bt]
    \centering
    \setlength{\tabcolsep}{7pt}
    \begin{tabular}{ccccc}
     \hline \hline
      & Material &  Edge length & Thickness  & Mass  \\ 
     \hline
     cube  & Tungsten  & 0.804  cm &  -           & 10.0 g  \\ 
     shell & Tungsten  & 4.166 cm &  50 $\mu$m  & 10.0 g   \\
     \hline \hline
    \end{tabular}    
    \caption{Parameters of the test bodies; their edge lengths ($L_{\mathrm{cube}}$\,\& $L_{\mathrm{shell}}$), the thickness of the shell ($\delta$), and their masses ($m_\mathrm{tot}$). The density of the tungsten is $19.25\,\mathrm{g/cm^3}$.}
    \label{tab:parameter}
\end{table}

\vspace{0.3cm}

\noindent
{\bf Projected Sensitivity \& Current Constraints} -- With these setup parameters, the scattering cross-sections of the DM off the cube (solid green curve) and the shell (solid blue curve) are shown in \gfig{fig:xsec}. Due to the difference in their edge lengths and form factors, there is a considerable difference in their cross-sections $|\sigma_{\mathrm{tot}}^{\mathrm{cube}} - \sigma_{\mathrm{tot}}^{\mathrm{shell}}|$, which is also shown in the figure as a dashed black curve.
This difference in cross-section can induce measurable differential accelerations detectable in the proposed experimental setup. 

Physical signatures must further incorporate the DM phase-space distribution.
The velocities of the DM particles follow the Maxwell-Boltzmann distribution in the galactic frame, which is transformed into the distribution in the laboratory frame as $f(\mathbf{v}_\chi)$\,\cite{Baxter:2021pqo}. The averaged acceleration of the test bodies induced by the DM scattering is obtained through integration, 
\begin{equation}
    \langle a_z \rangle =
    \frac{n_\chi}{m_\mathrm{tot}} 
    \int d^3 \mathbf{v}_\chi\,d\Omega'_\chi\,
    q_z\,|\mathbf{v}_\chi|\,
    f(\mathbf{v}_\chi)\,
    \frac{d\sigma_\mathrm{tot} (q)}{d \Omega'_\chi}.
\end{equation}
The integration element $d^3 \mathbf{v}_\chi$ contains the initial DM direction $\theta_\chi$ and $\phi_\chi$, while $d \Omega'_{\chi}$ represents the final DM solid angles $\theta'_\chi$ and $\phi'_\chi$. Due to the coherent form factor, the total cross-section is a function of the momentum transfer amplitude $q = m_\chi |\mathbf{v}_\chi - \mathbf{v}'_\chi|$. The torque is sensitive to acceleration along the direction perpendicular to the fiber, which we take as the $z$-axis\footnote{In general, the $z$-axis, which is aligned with the Earth's surface and dependent on the laboratory's latitude, forms an angle with the DM wind direction. In this analysis, we assume that DM is incident along the $z$-axis to estimate the maximal signal strength.}. Therefore, only momentum transfer in this direction, $q_z = m_\chi v_\chi (\cos \theta_\chi - \cos \theta'_\chi)$, contributes.
The detectable acceleration difference $\Delta a_z$ is obtained by performing the velocity average for both the cube and the shell, 
$\Delta a_z \equiv |\langle a_z \rangle_{\mathrm{cube}} - \langle a_z \rangle_{\mathrm{shell}}|$.

\begin{figure}[t]
    \centering
    \includegraphics[width=8.5cm]{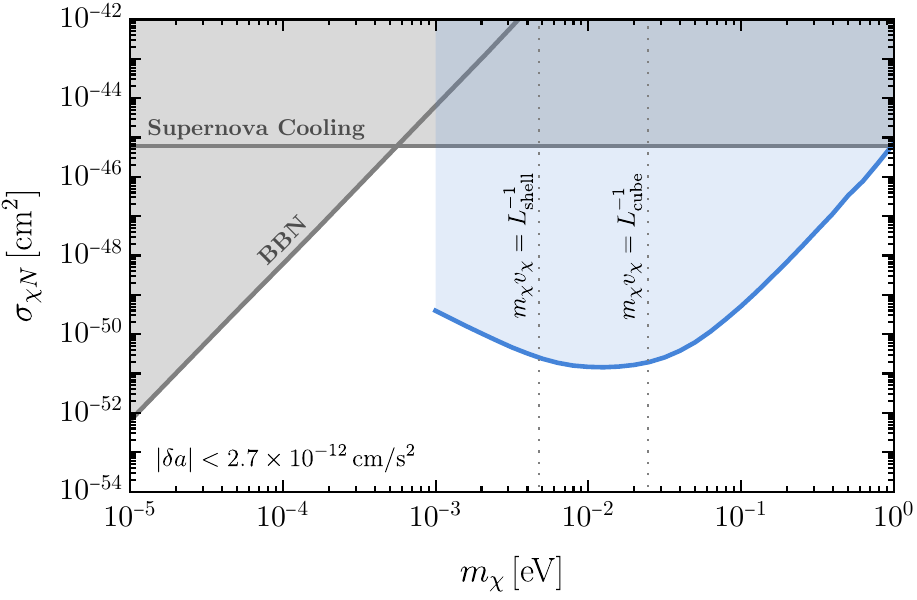}
    \caption{The projected \SI{95}{\percent} C.L. sensitivity to the DM-nucleon scattering cross-section $\sigma_{\chi N}$ as a function of the DM mass $m_\chi$ (shaded blue region). The present constraints on the light DM from supernova cooling and BBN are also shown (shaded grey regions). The typical DM velocity is $v_\chi = 10^{-3}$.}
\label{fig:limit}
\end{figure}

Within the sensitivity $|\delta a| < 2.7\times 10^{-12}\,\mathrm{cm/s^{2}}$ at \SI{95}{\percent} confidence level, the detectable parameter space is shown as the blue shaded area in \gfig{fig:limit}.
Depending on the DM mass and coherent-scattering length $\lambda \equiv 1/(m_\chi v_\chi) \simeq 10^3/ m_\chi$, the detectable cross-section behaves differently. As the acceleration is proportional to the difference between the total cross-sections, the sensitivities become stronger (weaker) in the mass regions where the cross-section difference increases (decreases), while the sensitivity is flat in the regions where the difference remains constant. The strongest sensitivity is obtained in the case that DM wavelength is between the dimension of cube and shell, $L_{\mathrm{shell}} > \lambda > L_{\mathrm{cube}}$. In this mass range, DM coherently scatters only with the cube while the cross-section of the shell drops sharply, creating a significant cross-section difference and the sensitivity reaches to $\sigma_{\chi N} \simeq 10^{-51}\,$cm$^2$.
The projected sensitivity stops at $m_\chi \sim 10^{-3}\,$eV, as lighter DM particles with de Broglie wavelengths exceeding the  characteristic scale of the torsion would coherent scatter with the entire apparatus without distinguishing the cube and shell.
Due to the small DM scattering cross-section constrained by our experimental setup, DM is not shielded by the Earth atmosphere or laboratory walls\,\cite{Day:2023mkb}.

Interestingly, the projected limit can be extended into the light fermion mass region, $(0.1,\,1)$\,keV\,\cite{Tremaine:1979we, Domcke:2014kla}, though the coherent effect starts decreasing, giving $\sigma_{\chi N} \lesssim (10^{-36},\,10^{-32})$\,cm$^2$. This is stronger than the constraints from the direct detection of such fermion DMs\,\cite{Bringmann:2018cvk, PROSPECT:2021awi, Afek:2021vjy}.

For a light bosonic DM that can scatter with a nucleon, the same interaction vertex also affects star cooling due to DM emission. The strongest constraint comes from the supernova SN1987A by requiring that the instantaneous luminosity of the DM emission should be lower than that of neutrinos\,\cite{Raffelt:1990yz, Olive:2007aj}. Taking the interaction $\mathcal L_\text{eff} = \bar n n \chi^2 / f_n + \bar p p \chi^2 / f_p$ with the same interaction strength for a neutron and a proton (i.e., $f_p$ = $f_n$) as an example, the DM cooling process imposes the constraint depicted by the grey shaded region in \gfig{fig:limit},
based on the limit established in Ref.~\cite{Olive:2007aj,Day:2023mkb}.
Moreover, such an interaction also alters the effective mass of neutrons and protons, and consequently affects the helium abundance in the process of Big Bang Nucleosynthesis (BBN)\,\cite{Bouley:2022eer,Day:2023mkb}. The specific impact depends on how the DM couples to quarks ($m_i \chi^2 \bar{\psi}_i \psi_i/ \Lambda^2$ with $i = u, d$) and gluons ($\chi^2 G^{\mu \nu} G_{\mu \nu} / \Lambda^2$) in a UV-complete model. Different coupling methods in the UV model do not yield a significant difference, so we illustrate the constraint from the BBN on $\sigma_{\chi N}$ as shown by the diagonal line in the figure, assuming that all the DM-nucleon interactions arise from the DM-gluon coupling. 
Additionally, DM-photon coupling can be loop-induced from DM interactions with gluons, quarks, or nucleons~\cite{Cox:2024rew,Ge:2024cto}, leading to DM thermalization in the early Universe. To minimize contributions to $N_\mathrm{eff}$ and avoid impacting BBN, DM should decouple before BBN, requiring $\sigma_{\chi N} \lesssim 10^{-35} \, \mathrm{cm}^2$.
DM-quark/gluon coupling can also contribute to meson decays. Constraints from $K^+$ decay on DM-quark interactions impose a bound of $\sigma_{\chi N} \lesssim 5 \times 10^{-39} \, \mathrm{cm}^2$~\cite{Cox:2024rew}.
Even after considering these astrophysical and cosmological constraints, the projected sensitivity from our proposed experiment surpasses those in the mass range of $(10^{-3},\,1)$\,eV.
\vspace{0.3cm}

\noindent
{\bf Discussion and Conclusions} -- The coherent scattering effect significantly enhances the scattering cross-section between a light DM and a macro-object, so the acceleration of the target induced by the DM scattering is amplified. We propose a torsion balance experiment with test bodies of distinct geometries --- a cube and a shell --- to detect this acceleration. The geometric disparity results in different scattering cross-sections with the DM, as well as measurable acceleration and torque. This design is able to provide the strongest experimental sensitivity on the DM scattering cross-section with a nucleus, $\sigma_{\chi N} \sim 10^{-51}\,$cm$^2$, for the mass range of the DM, $m_\chi \sim (10^{-3},\,1)$\,eV.

\section*{Acknowledgements}

\noindent
The authors thank Yu Cheng, Shao-Feng Ge, Maxim Khlopov, Yu-Cheng Qiu for fruitful discussions. P. Luo is supported by the National Natural Science Foundation of China (NSFC) (No.~12475052).
S. Matsumoto is supported by Grant-in-Aid for Scientific Research from MEXT, Japan; 23K20232\,(20H01895), 20H00153, 24H00244, 24H02244, by JSPS Core-to-Core Program; JPJSCCA20200002, and by World Premier International Research Center Initiative (WPI), MEXT, Japan (Kavli IPMU). J. Sheng is supported by the National Natural Science Foundation of China (Nos. 12375101, 12425506, 12090060, 12090064) and the SJTU Double First Class start-up fund WF220442604.
C.-Y. Xing is supported by the Fundamental Research Funds for the Central Universities (No.~24CX06048A). L. Zhu is supported by the National Natural Science Foundation of China (No. 12005308 and No. 12150012).

\providecommand{\href}[2]{#2}\begingroup\raggedright\endgroup

\end{document}